\documentclass[aps,pra,showpacs,twocolumn]{revtex4}

\usepackage{amsmath}
\usepackage{amssymb}
\usepackage{graphicx}% Include figure files
\usepackage{dcolumn}% Align table columns on decimal point
\usepackage{bm}% bold math

\newcommand{\ga}{\gamma}

\newcommand{\prt}{\partial}

\begin{document}

\title{
Collision of rarefaction waves in Bose-Einstein condensates}

\author{ S. K. Ivanov}
\affiliation{Moscow Institute of Physics and Technology, Institutsky
lane 9, Dolgoprudny, Moscow region, 141700, Russia}
\affiliation{Institute of Spectroscopy,
Russian Academy of Sciences, Troitsk, Moscow, 108840, Russia}

\author{ A. M. Kamchatnov}
\affiliation{Moscow Institute of Physics and Technology, Institutsky
lane 9, Dolgoprudny, Moscow region, 141700, Russia}
\affiliation{Institute of Spectroscopy,
Russian Academy of Sciences, Troitsk, Moscow, 108840, Russia}

%\date{\today}

\begin{abstract}
We consider the problem of expansion of Bose-Einstein condensate released form a box.
On the contrary to the standard situation of release from a harmonic trap, in this case
the dynamics is complicated by a process of collision of two rarefaction waves
propagating to the center of the initially uniform distribution. Complete analytical
solution of this problem is obtained by Riemann method in hydrodynamic dispersionless
approximation and the results are compared with the exact numerical solution of the
Gross-Pitaevskii equation.
\end{abstract}


\maketitle

\section{Introduction}
\label{intro}

One of the basic problems in dynamics of Bose-Einstein condensates (BECs) is that of
expansion after its release from a trap, because in many experimental situations measurements
are performed at the state of BEC's inertial expansion which properties are predetermined
to much extend by the initial stage of evolution. The simplest approach to this problem
was formulated and studied in Refs.~\cite{cd-96,kss-96} in hydrodynamic approximation for
the case of harmonic traps when the initial state of BEC is described accurately enough
by the Thomas-Fermi distribution. More detailed study of this problem with the use of
classical approach of Talanov \cite{talanov} was given in Refs.~\cite{bkk-03,kamch-04}.
These solutions were self-similar and at every moment of time the space distribution of
the density had the parabolic Thomas-Fermi time-dependent form. However, if the trap
is not harmonic, then the expansion is not self-similar anymore and some characteristic
features of the initial distribution can persist for quite long period of evolution
to be noticeable experimentally. For example, this happens after release of BEC
from a box-like trap with a uniform potential realized experimentally in
Refs.~\cite{gsgsh-13,gsgnsh-14} where the initial size $2l$ of a box plays the role
of the parameter which determines the expansion dynamics for time $t\sim l/c_0$,
where $c_0$ is the sound velocity at the initial uniform state of BEC. Indeed,
the evolution begins with propagation of two rarefaction waves from the edges of
BEC and these two waves collide at the center of the initial distribution at the
moment $t_c=l/c_0$ after which the distribution of the density acquires quite complicated
form different from parabolic self-similar distributions known from
Refs.~\cite{cd-96,kss-96,bkk-03,kamch-04}. The aim of this paper is to study such
an evolution in hydrodynamic approximation and to reveal its characteristic features.
To solve this problem analytically, we use the powerful Riemann method developed
in compressible fluid dynamics with quite general equation of state (see, e.g.,
\cite{sommer-49,ch-62}) which seems most suitable in BEC's hydrodynamics case
with its non-standard ``adiabatic index'' $\gamma=2$ (see, e.g., Ref.~\cite{wfml-99},
where the wave breaking problem was considered by this method in similar nonlinear
optics context).

\section{Formulation of the problem}

To demonstrate specific features of evolution of BEC expansion after its release from
a box-like trap, we consider one-dimensional situation where the dynamics is governed
by the Gross-Pitaevskii (GP) equation
\begin{equation}\label{eq1}
  i\psi_t+\frac12\psi_{xx}-|\psi|^2\psi=0,
\end{equation}
written here in standard non-dimensional variables. Transition from the BEC wave
function $\psi$ to more convenient in hydrodynamics variables density $\rho$ and
flow velocity $u$ is performed by means of the substitution
\begin{equation}\label{eq2}
    \psi(x,t)=\sqrt{\rho(x,t)}\exp\left({i}\int^x u(x',t)dx'\right),
\end{equation}
so that the GP equation is cast to the system
\begin{equation}\label{eq3}
\begin{split}
 &\rho_t+(\rho u)_x=0,\\
 &u_t+uu_x+\rho_x+\left[\frac{\rho_x^2}{8\rho^2}
   -\frac{ \rho_{xx}}{4\rho}\right]_x= 0.
\end{split}
\end{equation}
The last term in the second equation describes the dispersive effects and
in the hydrodynamic approximation it can be neglected since we consider
evolution of BEC cloud with mainly smooth enough dependence of $\rho$ and $u$
on the space coordinate $x$. As a result, we arrive at the so-called
``shallow water'' equations
\begin{equation}\label{eq4}
    \rho_t+(\rho u)_x=0,\quad u_t+uu_x+\rho_x=0.
\end{equation}
At the initial moment of time the distribution of density is uniform
within the interval $-l\leq x\leq l$,
\begin{equation}\label{eq5}
  \rho(x,0)=\left\{
  \begin{array}{ll}
  \rho_0,\qquad &|x|\leq l,\\
  0,\qquad &|x|>l.
    \end{array}
    \right.
\end{equation}
Although this distribution cannot be considered as ``smooth'', we shall show
later by comparison of hydrodynamic approximation with the exact numerical solution
of the GP equation (\ref{eq1}), that if $l\gg1$ (that if the size of the trap
is much greater than the healing length), then deviations of the exact solution
from its hydrodynamic approximation is negligibly small almost everywhere
except small regions at the boundaries of the BEC cloud with vacuum.

Since the Riemann method is not commonly used in theoretical physics, we shall
provide in the next section the relevant basic information about it.

\section{Riemann method}

For future convenience, we consider the compressible fluid dynamics equations
with adiabatic equation of state, $p=\rho^{\gamma}/\gamma$, where $p$ denotes
the pressure in the gas,
\begin{equation}\label{eq6}
    \rho_t+(\rho u)_x=0,\quad u_t+uu_x+\rho^{\gamma-2}\rho_x=0,
\end{equation}
so that Eqs.~(\ref{eq4}) are reproduced for $\gamma=2$. These equations can
be cast into diagonal Riemann form by introduction of new variables, namely Riemann
invariants
\begin{equation}\label{eq7}
    r_{\pm}=\frac{u}2\pm\frac1{\ga-1}\rho^{\frac{\ga-1}2},
\end{equation}
for which we get the equations
\begin{equation}\label{53-6}
\begin{split}
  & \frac{\prt r_+}{\prt t}+v_+(r_+,r_-)\frac{\prt r_+}{\prt x}=0,\\
  & \frac{\prt r_-}{\prt t}+v_-(r_+,r_-)\frac{\prt r_-}{\prt x}=0,
  \end{split}
\end{equation}
where
\begin{equation}\label{57-7}
\begin{split}
   & v_+=\frac12[(1+\ga)r_++(3-\ga)r_-],\\
   & v_-=\frac12[(3-\ga)r_++(1+\ga)r_-].
    \end{split}
\end{equation}

\begin{figure}[t]
\begin{center}
\includegraphics[width=8cm]{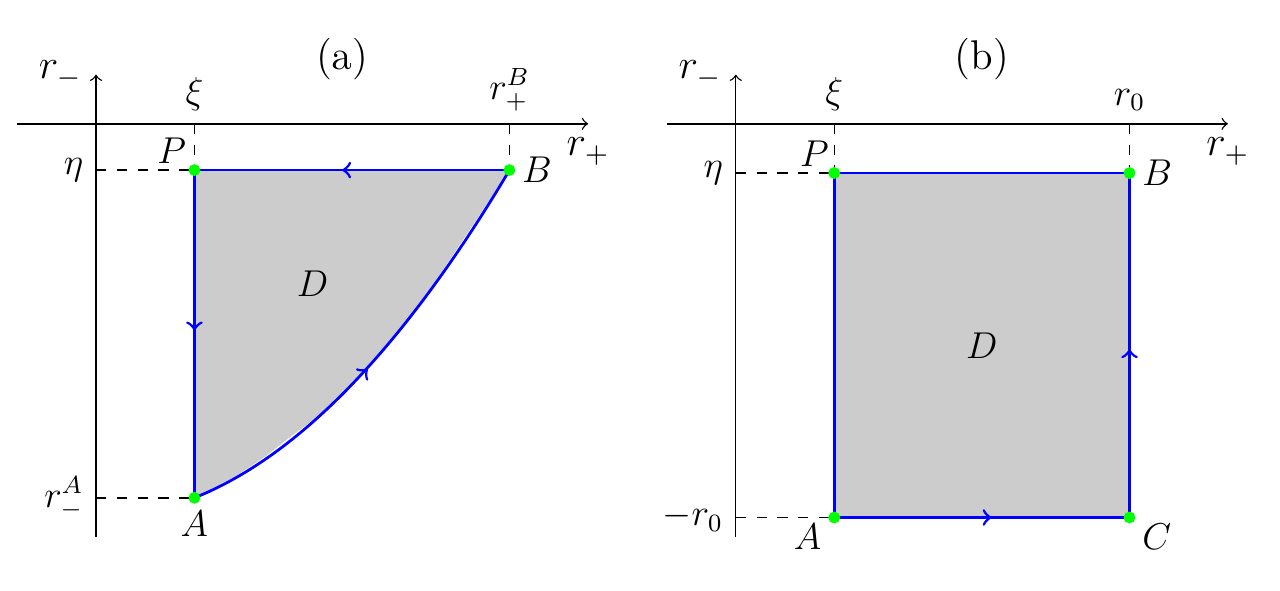}
\caption{(a) The initial data are given along the curve $AB$ in the hodograph plane in the general
formulation of the Riemann method. (b) The segments $AC$ and $CB$ form the ``initial data curve''
for the problem of collision of two rarefaction waves by the Riemann method.
}
\label{fig1}
\end{center}
\end{figure}

Riemann noticed that Eqs.~(\ref{53-6}) become linear with respect to
the dependent variables if one considers $x$ and $t$ as functions of
the Riemann invariants, $x=x(r_+,r_-),\,t=t(r_+,r_-)$, and after this
``hodograph transform'' we arrive at the system
\begin{equation}\label{56-5}
\begin{split}
    & \frac{\prt x}{\prt r_-}-v_+(r_+,r_-)\frac{\prt t}{\prt r_-}=0,\\
    & \frac{\prt x}{\prt r_+}-v_-(r_+,r_-)\frac{\prt t}{\prt r_+}=0.
    \end{split}
\end{equation}
We look for the solution of this system in the form
\begin{equation}\label{57-1}
    \begin{split}
    x-v_+(r_+,r_-)t=w_+(r_+,r_-),\\
    x-v_-(r_+,r_-)t=w_-(r_+,r_-).
    \end{split}
\end{equation}
Their substitution into Eqs.~(\ref{56-5}) and elimination of $t$ yields
with account of Eqs.~(\ref{57-7})
\begin{equation}\nonumber
\begin{split}
    \frac1{w_+-w_-}\frac{\prt w_+}{\prt r_-}=
    \frac1{v_+-v_-}\frac{\prt v_+}{\prt r_-}=\frac{\beta}{r_+-r_-},\\
    \frac1{w_+-w_-}\frac{\prt w_-}{\prt r_+}=
    \frac1{v_+-v_-}\frac{\prt v_-}{\prt r_+}=\frac{\beta}{r_+-r_-},
    \end{split}
\end{equation}
where
\begin{equation}\label{58-1}
    \beta=\frac{3-\ga}{2(\ga-1)}.
\end{equation}
This means that $\prt w_+/\prt r_-=\prt w_-/\prt r_+$ and, hence,
we can represent $w_{\pm}$ as
\begin{equation}\label{58-2}
    w_+=\frac{\prt W}{\prt r_+},\quad w_-=\frac{\prt W}{\prt r_-},
\end{equation}
where $W$ is a solution of the Euler-Poisson (EP) equation
\begin{equation}\label{58-3}
    \frac{\prt^2W}{\prt r_+\prt r_-}-
    \frac{\beta}{r_+-r_-}\left(\frac{\prt W}{\prt r_+}-\frac{\prt W}{\prt r_-}\right)=0.
\end{equation}
The characteristics of this second order partial differential equation are
the straight lines $r_+=\xi=\mathrm{const}$, $r_-=\eta=\mathrm{const}$ parallel to the
coordinates axes in the hodograph plane. The Riemann method is based on the idea that
one can find the solution of the EP equation in the form similar to d'Alembert solution
of the wave equation with explicit account of the initial conditions given
on some curve $AB$ in the hodograph plane (see Fig.~\ref{fig1}(a)). These data are
transferred along the characteristics into the domain of dependence $D$, so that
the function $W$ can be found at any point $P(\xi,\eta)\in D$.

Riemann showed (see, e.g., Refs.~\cite{sommer-49,ch-62}) that $W(P)$ can be
represented in the form
\begin{equation}\label{m1-282.9}
  W(P)=\frac12(R\overline{W})_A+\frac12(R\overline{W})_B+\int_A^B(Vdr_++Udr_-),
\end{equation}
where
\begin{equation}\label{m1-284.4}
\begin{split}
 & U=\frac12 \left(R\frac{\prt \overline{W}}{\prt r_-}-
 \overline{W}\frac{\prt R}{\prt r_-}\right)-\frac{\beta}{r_+-r_-}\overline{W}R,\\
 & V=\frac12 \left(\overline{W}\frac{\prt R}{\prt r_+}
 -R\frac{\prt \overline{W}}{\prt r_+}\right)-\frac{\beta}{r_+-r_-}\overline{W}R,
 \end{split}
\end{equation}
$\overline{W}$ in the right-hand sides represents values of $W$ along the boundary
arc $AB$ in the hodograph plane (see Fig.~\ref{fig1}(a)), and $R$ if the Riemann
function which satisfies the equation
\begin{equation}\label{eq16b}
  \frac{\prt^2R}{\prt r_+\prt r_-}+\frac{\beta}{r_+-r_-}\left(\frac{\prt R}{\prt r_+}-
  \frac{\prt R}{\prt r_-}\right)-\frac{2\beta R}{(r_+-r_-)^2}=0
\end{equation}
and its solution case can be expressed in term of hypergeometric function $F(a,b;c;z)$
(see \cite{sommer-49})
\begin{equation}\label{m1-284.2}
\begin{split}
  & R=\left(\frac{r_+-r_-}{\xi-\eta}\right)^{\beta}F(\beta,1-\beta;1;z),\\
  & z=\frac{(r_+-\xi)(r_--\eta)}{(r_+-r_-)(\xi-\eta)},
  \end{split}
\end{equation}
where $\xi$ and $\eta$ are the coordinates of the point $P$ in the hodograph plane $(r_+,r_-)$.
Now we can turn to our problem of collision of two rarefaction waves in BEC.

\section{Collision of rarefaction waves}

In BEC we have $\ga=2$ and, consequently, $\beta=1/2$, so Eqs.~(\ref{53-6}) for the Riemann
invariants $r_{\pm}=u/2\pm\sqrt{\rho}$ take the form
\begin{equation}\label{eq18}
\begin{split}
  & \frac{\prt r_+}{\prt t}+\frac12(3r_++r_-)\frac{\prt r_+}{\prt x}=0,\\
  & \frac{\prt r_-}{\prt t}+\frac12(r_++3r_-)\frac{\prt r_-}{\prt x}=0.
  \end{split}
\end{equation}
Before the moment of the collision, i.e., for for $t<l/c_0$, $c_0=\sqrt{\rho_0}$,
the rarefaction waves are given by the simple wave solutions of the system (\ref{eq18}),
\begin{equation}\label{eq19}
\begin{split}
  r_+=c_0,\quad & x-\frac12(c_0+3r_-)t=\frac{\prt W}{\prt r_-}=l,\\
  & l-c_0t\leq x\leq l+2c_0t;\\
  r_-=-c_0,\quad & x-\frac12(3r_+-c_0)t=\frac{\prt W}{\prt r_+}=-l,\\
  & -l-2c_0t\leq x\leq -l+c_0t,
\end{split}
\end{equation}
and the condensate remains at rest with the density $\rho_0$ in the region
$-l+c_0t\leq x\leq l-c_0t$. After the moment $l/c_0$ the region of the general
solution of the system (\ref{eq18}) appears in the interval $x_L(t)\leq x\leq x_R(t)$
where  both Riemann invariants change with time and space coordinate, and this general
solution matches with the simple waves (\ref{eq19}) at the points $x_L(t)$ and $x_R(t)$.
This means that in the hodograph plane the function $W$ must satisfy the boundary
conditions
\begin{equation}\label{eq20}
\begin{split}
  & \frac{\prt W}{\prt r_-}=l\quad\text{at}\quad r_+=c_0,\\
  & \frac{\prt W}{\prt r_+}=-l\quad\text{at}\quad r_-=-c_0,
  \end{split}
\end{equation}
that is \begin{equation}\label{eq21}
  \overline{W}=-l(r_+-r_-)
\end{equation}
on the sides $AC$ and $CB$ of the rectangle $ACBP$ shown in Fig.~\ref{fig1}(b).
If we solve the EP equation (\ref{58-3}) ($\beta=1/2$) with this
boundary condition, then we can find $W$ at the point $P(\xi,\eta)$,
and then the values $r_+=\xi, r_-=\eta$ of the Riemann invariants at this
point are related with $x$ and $t$ by the formulae (\ref{57-1}), that is
\begin{equation}\label{eq22}
    \begin{split}
    x-\frac12(3\xi+\eta)t=\frac{\prt W}{\prt \xi},\quad
    x-\frac12(\xi+3\eta)t=\frac{\prt W}{\prt \eta}.
    \end{split}
\end{equation}

\begin{figure}[t]
\begin{center}
\includegraphics[width=8cm]{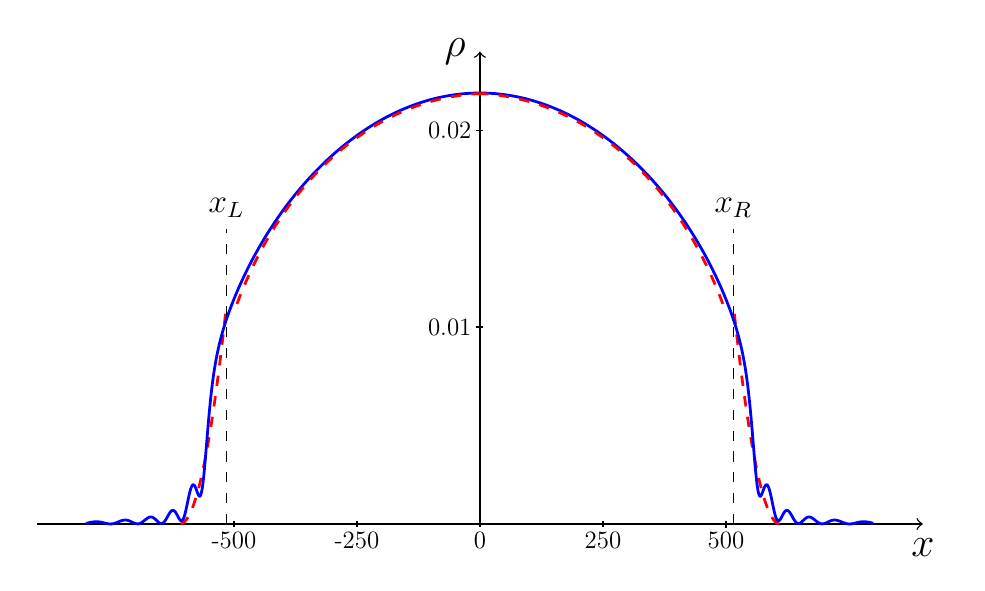}
\caption{Distribution of the density $\rho$ at the moment $t=300$. At the initial moment
it is given by Eq.~(\ref{eq5}) with $\rho_0=1$, $l=10$. Solid (blue) line shows the numerical
solution of the Gross-Pitaevskii equations and dashed (red) line corresponds to the
hydrodynamic approximation. Symbols $x_R$ and $x_L$ at vertical dashed lines indicate
the boundaries between the general solution and the simple waves.
}
\label{fig2}
\end{center}
\end{figure}

To find the solution of Eq.~(\ref{58-3}), we use the Riemann formula
(\ref{m1-282.9}) where for $\beta=1/2$ the Riemann function (\ref{m1-284.2})
can be transformed to more convenient expression
\begin{equation}\label{eq23}
\begin{split}
  & R=\frac2{\pi}\cdot\frac{r_+-r_-}{\sqrt{(r_+-\eta)(\xi-r_-)}}K(m),\\
  & m=\frac{(r_+-\xi)(\eta-r_-)}{(r_+-\eta)(\xi-r_-)},\qquad 0\leq m\leq 1,
  \end{split}
\end{equation}
where $K(m)$ is the elliptic integral of the first kind. Substitution of
Eq.~(\ref{eq21}) into Eq.~(\ref{m1-282.9})
followed by integration by parts with account of (\ref{m1-284.4}) yields
\begin{equation}\label{eq24}
  W(P)=(RW)_C+\frac{3l}2\left\{\int_{\xi}^{r_0}Rdr_+ +\int_{-r_0}^{\eta} Rdr_-\right\},
\end{equation}
where the first integral is taken along the side $AC$ of the rectangle in
Fig.~\ref{fig1}(b), the second one along the side $CB$, and at the point $C$ we have
$r_+=c_0,\,r_-=-c_0,\,m=m_0$,
\begin{equation}\label{eq25}
  m_0=\frac{(c_0-\xi)(c_0+\eta)}{(c_0+\xi)(c_0-\eta)}.
\end{equation}
Changing integration over $r_+$ and $r_-$ to integration over corresponding
specification of the variable $m$ yields the final expression
\begin{equation}\label{eq26}
\begin{split}
  W(\xi,\eta)=&-\frac{8lc_0^2}{\pi}\frac{K(m_0)}{\sqrt{(c_0+\xi)(c_0-\eta)}}\\
  & +\frac{3l}{\pi}\sqrt{\xi-\eta}\int_0^{m_0}F(\xi,\eta,m)dm,
  \end{split}
\end{equation}
where
\begin{equation}\label{eq27}
\begin{split}
  &F(\xi,\eta,m)=\Big\{\frac{(r_0+\xi)^{3/2}(r_0+\eta)^{3/2}}{[r_0+\eta-(r_0+\xi)m]^{5/2}}\\
  & + \frac{(r_0-\xi)^{3/2}(r_0-\eta)^{3/2}}{[r_0-\xi-(r_0-\eta)m]^{5/2}}\Big\}(1-m)K(m).
  \end{split}
\end{equation}
These formulae together with Eqs.~(\ref{eq22}) define implicitly the Riemann
invariants as functions of $x$ and $t$ and, consequently, the values of the density and
the flow velocity,
\begin{equation}\label{eq28}
  \rho=\frac14(\xi-\eta)^2,\qquad u=\xi+\eta.
\end{equation}
We compare in Fig.~\ref{fig2} the analytical results (dashed thick lines) with the exact
numerical solution (solid line) corresponding to the initial condition (\ref{eq5})
with $\rho_0=1$, $l=10$, and the evolution time $t=300$. As one can see, the
agreement is very good almost everywhere except for the edges of the wave near the
boundaries with vacuum where the regions of small oscillations appear. Such
oscillations are generated due to dispersion effects and they originate from the
sharp dependence of the initial distribution (\ref{eq5}) of the density on $x$ at
the edges $x=\pm l$. Thus, the hydrodynamic approximation gives accurate enough
description of the wave resulting from collision of two rarefaction waves in BEC.

Although the above formulae provide the complete solution of our problem, its analysis
can be considerably simplified by the following remark.
The compatibility condition $\prt^2x/\prt \xi\prt \eta= \prt^2x/\prt \eta\prt \xi$ of
the equations (see Eqs.~(\ref{56-5}))
\begin{equation}\label{t-53.51}
  \frac{\prt x}{\prt \xi}-\frac12(\xi+3\eta)\frac{\prt t}{\prt \xi}=0,\qquad
  \frac{\prt x}{\prt \eta}-\frac12(3\xi+\eta)\frac{\prt t}{\prt \eta}=0.
\end{equation}
yields the Euler-Poisson equation for the function $t=t(\xi,\eta)$,
\begin{equation}\label{t-53.53}
  \frac{\prt^2t}{\prt \xi\prt \eta}-
\frac{3}{2(\xi-\eta)}\left(\frac{\prt t}{\prt \xi}-\frac{\prt t}{\prt \eta}\right)=0.
\end{equation}
Its solution satisfying the necessary boundary conditions can be found by
the same method as Eq.~(\ref{eq16b}) is solved (see \cite{sommer-49})
and it is given by the formula
\begin{equation}\label{t-53.54}
  t=\frac{8lc_0^2}{(c_0+\xi)^{3/2}(c_0-\eta)^{3/2}}\,
  F\left[\frac32,\frac32;1;\frac{(c_0-\xi)(c_0+\eta)}{(c_0+\xi)(c_0-\eta)}\right],
\end{equation}
where $F$ is again the hypergeometric function. This formula gives the dependence of time
$t$ on $\xi$ and $\eta$ in the whole region of the general solution.

\begin{figure}[t]
\begin{center}
\includegraphics[width=8cm]{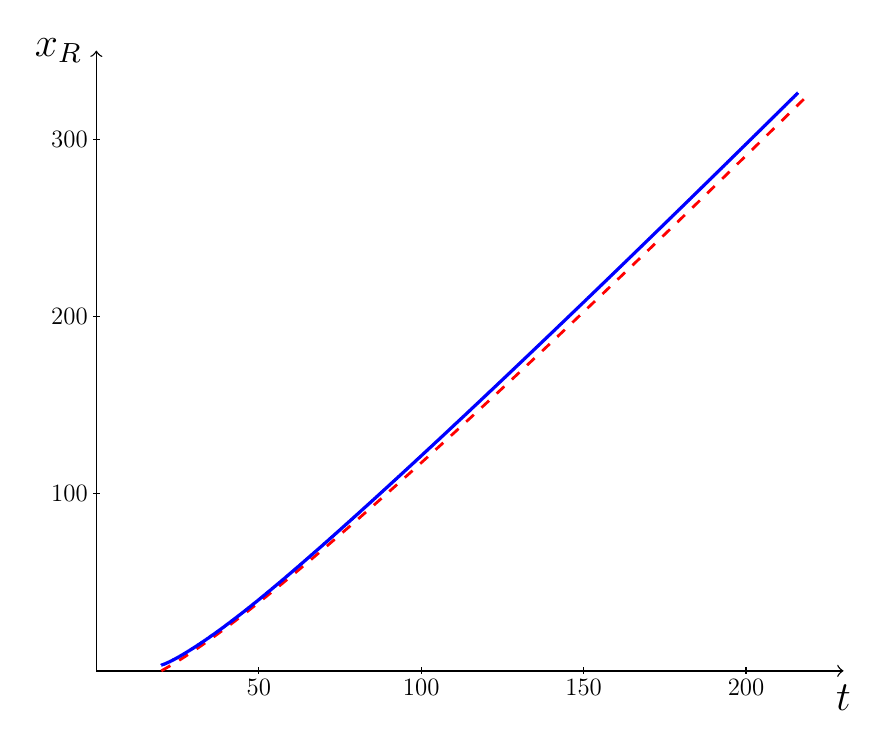}
\caption{Trajectory of the boundary $x_R(t)$ between the general solution and the right
simple wave. Solid blue line shows the numerical solution and dashed red line depicts
the plot of analytical formula (\ref{tt2-48.2}).
}
\label{fig3}
\end{center}
\end{figure}

At the right boundary between the general solution and the rarefaction wave we have
$\xi=c_0$, hence Eq.~(\ref{t-53.54}) simplifies to
\begin{equation}\label{t-53.55}
 t=\frac{2\sqrt{2}lc_0^{1/2}}{(c_0-\eta)^{3/2}},
\end{equation}
and elimination of $\eta$ from this equation and the first formula (\ref{eq19}) (with $r_-=\eta$)
for the rarefaction wave gives
the law of motion of this boundary,
\begin{equation}\label{tt2-48.2}
    x_R(t)=l+2c_0t-3l\left(\frac{c_0t}{l}\right)^{1/3}.
\end{equation}
Fig.~\ref{fig3} demonstrates good agreement of this analytical formula with
the numerical results.

At the center of the distribution we have $\eta=-\xi$ so that Eq.~(\ref{t-53.54}) yields
in implicit form the dependence of $\xi$ on $t$,
\begin{equation}\label{t-54.56}
  t=\frac{8lc_0^2}{(c_0+\xi)^{3}}\,
  F\left[\frac32,\frac32;1;\left(\frac{c_0-\xi}{c_0+\xi}\right)^2\right].
\end{equation}
Since the flow velocity $u$ vanishes here and, consequently, $\rho=\xi^2$, this
formula gives the dependence of the density $\rho$ on time $t$ at the center of
the wave at $x=0$. For asymptotically large time $t\gg l/c_0$ we get
with logarithmic accuracy
\begin{equation}\label{t3-136.21}
\begin{split}
  &\rho\approx\rho_0\Big\{\frac2{\pi}\cdot\frac{l}{c_0(t-t_0)}\\
  &+ \frac1{\pi^2}\cdot \left(\frac{l}{c_0(t-t_0)}\right)^2\ln\frac{c_0t}l\Big\},\\
 & t_0=\frac{l}{2\pi c_0}(7-4\gamma-5\ln 2-4\psi(3/2)\\
 &+\ln\pi+\psi(3/2))
   \approx 0.353988({l}/{c_0}),
\end{split}
\end{equation}
$\ga \approx 0.577216$ is the Euler constant, $\psi(z) = \Gamma'(z)/\Gamma(z)$.
Even the first term here give very good approximation to the exact expression
(\ref{t-54.56}) in the whole region $t>1$.

\begin{figure}[t]
\begin{center}
\includegraphics[width=8cm]{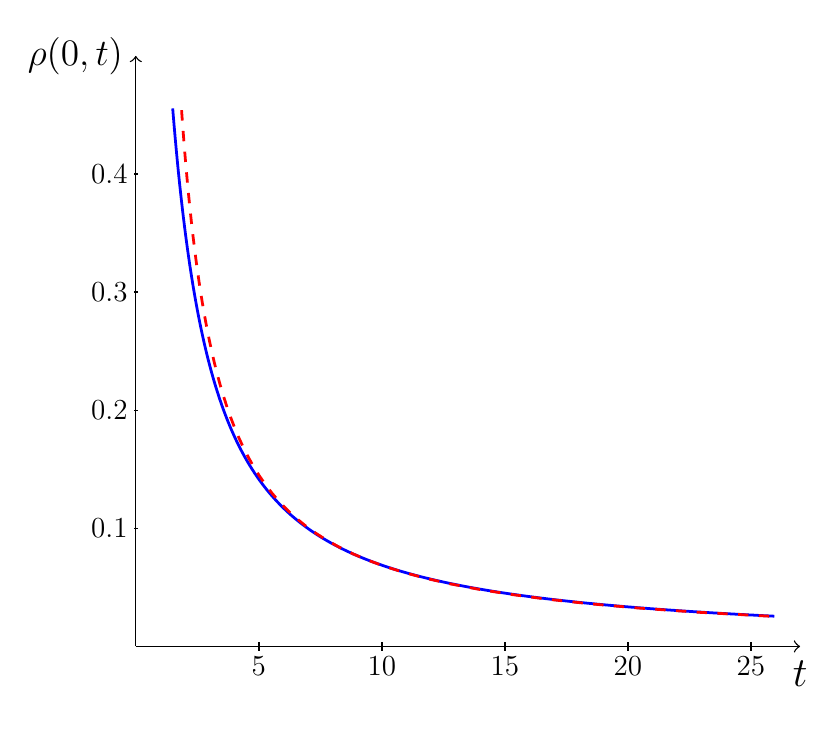}
\caption{Dependence of the density $\rho$ of BEC at the center of the wave
at $x=0$ on time $t$. The initial parameters are equal to $\rho_0=1$, $l=1$.
Exact numerical solution is shown by a solid blue line and analytical
approximation by a red dashed line.
}
\label{fig4}
\end{center}
\end{figure}

The analytical dependence $\rho(0,t)$ obtained in the dispersionless approximation
is compared with numerical solution of the GP equation in Fig.~\ref{fig4}.
We have chosen here the initial length of the distribution equal to $l=1$
to show that there is some difference between the
analytical theory and the exact solution at the initial stage of evolution, if the
initial size of the box is about one healing length. For larger time $t\gtrsim 5$
this difference disappears and if $l\sim 10$ it is negligibly small for all
values of time $t>l/c_0$.

\section{Conclusion}

Thus, the considered
in this paper problem demonstrates that the process of free expansion of BEC
released from a trap can be more complicated than such an expansion in the
case of harmonic potential traps, and the solution obtained here provides the
method of analytical description of such a process.
For scales greater than the healing length, the hydrodynamic dispersionless
approximation to the GP equation is a very convenient tool
for analytical investigations for the following reasons. First, we have in
our disposal a very well developed apparatus of the compressible gas dynamics
which can be successfully applied to concrete problems, as it is demonstrated
in this paper for the problem of collision of two rarefaction waves. Second, the
analytical solution provides the main characteristic parameters of the wave
as, for example, the size of the BEC cloud or its density at the center,
at any moment of time, what may be useful for quantitative estimates and
comparison with experiment. At last, the dispersionless solution can be part
of more complicated wave structures as it happens, for example, in experiments
with formation of dispersive shock waves in BEC \cite{hoefer-06} or in
similar experiments in nonlinear optics \cite{wjf-07}.


\begin{thebibliography}{99}


\bibitem{cd-96} Y. Custin and R. Dum, Phys. Rev. Lett., {\bf 77}, 5315 (1996).

\bibitem{kss-96}  Yu. Kagan, E. L. Surkov, and G. V. Shlyapnikov, Phys.
Rev. A {\bf 54}, R1753 (1996).

\bibitem{talanov} V. I. Talanov, Pis’ma Zh. Eksp. Teor. Fiz. {\bf 2}, 218 (1965)
[JETP Lett. {\bf 2}, 138 (1965)].

\bibitem{bkk-03} V. A. Brazhnyi, A. M. Kamchatnov, and V. V. Konotop,
Phys. Rev., {\bf 68,} 035603 (2003).

\bibitem{kamch-04} A. M. Kamchatnov, Zh. Eksp. Teor. Fiz., {\bf 125}, 1041 (2004)
[JETP, {\bf 98}, 908 (2004).

\bibitem{gsgsh-13} A. L. Gaunt, T. F. Schmidutz, I. Gotlibovych, R. P. Smith, and Z. Hadzibabic,
Phys. Rev. Lett., {\bf 110}, 200406 (2013).

\bibitem{gsgnsh-14} I. Gotlibovych, T. F. Schmidutz, A. L. Gaunt, N. Navon,
R. P. Smith, and Z. Hadzibabic, Phys. Rev., A {\bf 89}, 061604(R) (2014).

\bibitem{sommer-49} A. Sommerfeld, Partial Differential Equations in Physics, Academic Press, N. Y., 1949.

\bibitem{ch-62} R. Courant and D. Hilbert, Methods of Mathematical Physics, Vol. II,
Interscience Publishers, N. Y., 1962.

\bibitem{wfml-99} O. C. Wright, M. G. Forest, K. T.-R. McLaughlin,
%On the exact solution of the geometric optics approximation of the defocusing nonlinear
%Schr\"odinger equation,
Phys. Lett., A {\bf 257,} 170 (1999).

\bibitem{hoefer-06} M. A. Hoefer, M. J. Ablowitz, I. Coddington, E. A. Cornell, P. Engels, V. Schweikhard,
%Dispersive and classical shock waves in Bose-Einstein condensates and gas dynamics,
Phys. Rev. A {\bf 74}, 023623 (2006)

\bibitem{wjf-07} W. Wan, S. Jia, J. W. Fleischer,
%Dispersive superfluid-like shock waves in nonlinear optics,
Nat. Phys. {\bf 3}, (1) 46 (2007).

\end{thebibliography}
\end{document}